\documentclass{article}

\usepackage{tccml_workshop_iclr, times}

\usepackage[utf8]{inputenc} 
\usepackage[T1]{fontenc}    
\usepackage{hyperref}       
\usepackage{url}            
\usepackage{booktabs}       
\usepackage{amsfonts}       
\usepackage{nicefrac}       
\usepackage{microtype}      
\usepackage{multirow}       
\usepackage{graphicx}       
\usepackage{subcaption}     

\setcitestyle{numbers}

\title{Forecasting Tropical Cyclones with Cascaded Diffusion Models}

\author{%
  Pritthijit Nath \\
  Department of Computing \\
  Imperial College London \\ 
  \texttt{pritthijit.nath22@imperial.ac.uk}
  \And
  Pancham Shukla \\
  Department of Computing \\
  Imperial College London \\   \texttt{panchamkumar.shukla@imperial.ac.uk}
  \And
  Shuai Wang \\
  Department of Geography and Spatial Sciences \\
  University of Delaware \\   \texttt{shwang@udel.edu}
  \And
  César Quilodrán-Casas \\
  Department of Earth Science and Engineering\\
  Imperial College London \\ 
  \texttt{c.quilodran@imperial.ac.uk}
}

\iclrfinalcopy

\begin{document}
\setcitestyle{square}

\maketitle

\begin{abstract}
As tropical cyclones become more intense due to climate change, the rise of Al-based modelling provides a more affordable and accessible approach compared to traditional methods based on mathematical models. This work leverages generative diffusion models to forecast cyclone trajectories and precipitation patterns by integrating satellite imaging, remote sensing, and atmospheric data. It employs a cascaded approach that incorporates three main tasks: forecasting, super-resolution, and precipitation modelling. The training dataset includes 51 cyclones from six major tropical cyclone basins from January 2019 - March 2023. Experiments demonstrate that the final forecasts from the cascaded models show accurate predictions up to a 36-hour rollout, with excellent Structural Similarity (SSIM) and Peak-Signal-To-Noise Ratio (PSNR) values exceeding 0.5 and 20 dB, respectively, for all three tasks. The 36-hour forecasts can be produced in as little as 30 mins on a single Nvidia A30/RTX 2080 Ti. This work also highlights the promising efficiency of Al methods such as diffusion models for high-performance needs in weather forecasting, such as tropical cyclone forecasting, while remaining computationally affordable, making them ideal for highly vulnerable regions with critical forecasting needs and financial limitations. Code accessible at %
\url{https://github.com/nathzi1505/forecast-diffmodels}.
\end{abstract}

\section{Introduction}
Climate change is a pressing global issue causing unprecedented changes in the Earth's climate system, resulting in altered precipitation patterns and a surge in extreme rainfall events with devastating environmental consequences~\cite{IPCC.2021}. Rising global temperatures and changing atmospheric circulation patterns are significant drivers of these extreme events~\cite{Donat.2016}, posing challenges for water resource management, infrastructure planning, and disaster risk reduction~\cite{VanAalst.2006}. Advanced machine learning (ML) techniques have emerged as a promising solution for predicting and understanding extreme rainfall behaviour under climate change~\cite{Pouyanfar.2018}. These algorithms can analyse large datasets, capture complex spatio-temporal relationships, and make precise predictions without the need for explicit programming. Leveraging modern computing systems like GPUs and distributed architectures, ML offers a revolutionary approach to meteorological modelling, replacing traditional supercomputer-based simulations~\cite{Schalkwijk.2015}.

In recent times, diffusion models~\cite{Ho.2020} have garnered substantial attention across various domains, including weather forecasting, climate modelling, and image processing. Leinonen et al.~\cite{Leinonen.2023} introduced a latent diffusion model (LDM) for precipitation nowcasting, surpassing traditional methods and deep generative models in accuracy and uncertainty quantification. Bassetti et al.~\cite{Bassetti.2023} demonstrated the efficiency of diffusion models, particularly DiffESM, in emulating Earth System Models (ESMs) for analyzing extreme weather events while demanding fewer computational resources. Hatanaka et al. ~\cite{Hatanaka.2023} improved uncertainty modelling in weather and climate forecasts using score-based diffusion models, benefiting high-dimensional solar irradiance predictions. Addison et al.~\cite{Addison.2022} showcased diffusion models' ability to generate realistic high-resolution rainfall samples from low-resolution simulations for flood modelling. 

This study draws inspiration from previous works in atmospheric modelling that have utilised diffusion models.  It explores a novel approach by employing multiple diffusion models organized in a cascading manner to predict tropical cyclones (TC), which are a critical example of extreme rainfall events that have been intensified by climate change. In particular, this research contributes to the existing literature on AI-based atmospheric modelling in the following ways:
\begin{enumerate}
\item A novel adaptation of a framework utilising diffusion models (often used in natural language processing (NLP) tasks such as text-to-image generation, image-to-image transformation, etc.) in forecasting TCs using publicly available atmospheric reanalysis data such as ERA5~\cite{era5}~\cite{Lavers.2022} and IR 10.8{\textmu m} satellite data.
\item Experimental demonstration of highly effective model predictive ability having single-step SSIM and PSNR above 0.5 and 20 dB respectively, and strong accuracy with rollout around 36 hours.
\item Model design optimised on single GPUs (eg. Nvidia RTX 2080 Ti) to underscore the affordability argument of AI-based methods whilst exhibiting comparable predictive capabilities to simulation-based conventional methods utilising expensive supercomputers.
\end{enumerate}

\section{Data}
\subsection{Data Acquisition}

\begin{enumerate}
\item \textbf{Satellite Data:}~Infrared (IR) 10.8{\textmu m} for a total of 51 cyclones (above 2 in the Saffir-Simpson Hurricane Wind Scale~\cite{Allaby.2008}) that have been reported to have major landfall impact are extracted from six major basins as shown in Table \ref{tbl:cyclone-basins} over the time period between January 2019 to March 2023. 

\begin{table}[!h]
\small
\centering
\vspace{-0.5em}
\caption{List of TC basins along with their satellite data providers and cyclone counts}
\label{tbl:cyclone-basins}
\begin{tabular}{@{}lllll@{}}
\toprule
\textbf{Hemisphere} & \textbf{Basin} & \textbf{Satellite} & \textbf{Sub-Region} & \textbf{Count} \\ \midrule
\multirow{4}{*}{Northern} & North Indian Ocean & INSAT - 3D & South Asia & 13 \\
 & North Atlantic Ocean & GOES - East & US East Coast & 11 \\
 & Eastern Pacific Ocean & GOES - East/West & US West Coast & 4\\
 & Western Pacific Ocean & Himawari 8/9 & Philippines & 9\\ \midrule
\multirow{2}{*}{Southern} & South-West Indian Ocean & Meteosat - 9 & Madagascar & 8 \\
 & Australia & Himawari 8/9 & North Australia & 6\\ \midrule
\end{tabular}
\vspace{-1em}
\end{table}

\item \textbf{Atmospheric Data:}~Hourly ERA5 reanalysis data for four atmospheric variables as shown in Table ~\ref{tbl:era5-data} over the period from formation to dissipation is acquired from the Copernicus Climate Data Store for each recorded cyclone.

\end{enumerate}

\subsection{Data Processing}

\begin{enumerate}
\item \textbf{Bounding Box Formulation:}~Square bounding boxes are empirically determined for all 51 cyclones, to cover all regions of interest over the entire trajectory of a hurricane from formation to dissipation. These bounding boxes are then used to crop full disk IR 10.8{\textmu m} satellite images and the corresponding ERA5 data to generate the cyclone dataset.

\item \textbf{Metadata Creation:}~To facilitate easy cyclone data retrieval and address path accessibility
issues, a functional data structure is designed to contain all the required metadata to load a given cyclone from memory. Fields included are:- region, name, bounding box coordinates, ERA5 filenames, total filecount and a list of sub-data structures, each containing data fields for each satellite file such as date, filepath and the corresponding ERA5 index.

\item \textbf{Train Test Bifurcation:}~For a fair evaluation of the diffusion models used, 20\% of the entire cyclone dataset (calculated region-wise) is reserved
for test set evaluation purposes and the remaining 80\% of the dataset is used for training purposes.

\item \textbf{Dataloader Generation:}~ Task specific dataloaders are created on the metadata data structure to facilitate model training and streamlining the data loading process. First, raw satellite and ERA5 image data are downsized to 64x64 (forecasting and precipitation modelling) and 128x128 (super-resolution) and then divided into randomly sampled batches of a specified batch size. In addition, supplementary methods such as min-max normalisation and data augmentation such as rotate90 are introduced at the dataloader level to aid model training.

\end{enumerate}

\section{Methodology}
\subsection{Cascaded Structure}

\begin{figure}
\centering
\includegraphics[width=0.85\hsize]{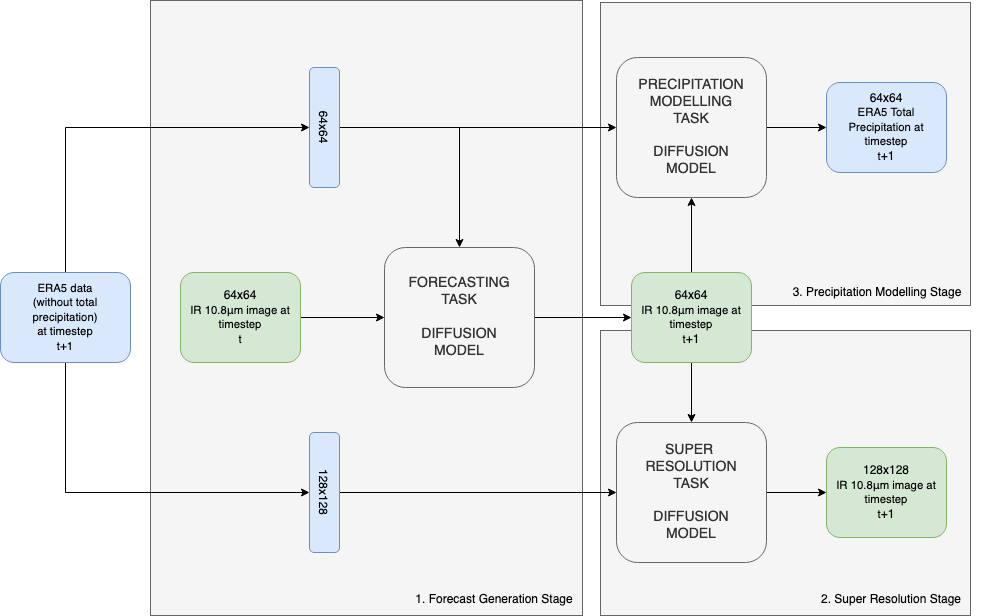}
\caption{Illustration of the cascaded arrangement involving three task-specific diffusion models}
\label{fig:cascaded-model}
\vspace{-1em}
\end{figure}

Taking inspiration from the Imagen paper~\cite{Saharia.2022} and its application in image generation conditioned on text inputs~\cite{lucidrans-imagen}, this study employs a cascaded arrangement, as depicted in Fig.~\ref{fig:cascaded-model}. In particular, this specific arrangement utilises three independently trained U-Net based diffusion models, each tailored to a specific task which ultimately enhances the efficiency of cyclone forecast generation. Using the 64x64 satellite IR 10.8{\textmu m} at time $t$, forecast at time $t+1$, is generated and pushed downstream onto the super-resolution task and the precipitation modelling task models. The super-resolution task model creates the 128x128 satellite 10.8{\textmu m} version of the generated 64x64 forecast, while the precipitation modelling task model generates the 64x64 total precipitation map corresponding to the forecast. In all three tasks, the forecasted ERA5 data at $t+1$ are used to condition the input.

For each diffusion model, a similar U-Net structure to Imagen~\cite{lucidrans-imagen} is used with additional refinements including classifier free guidance~\cite{Ho.2022}, dynamic thresholding (for maintaining the outputs within the normalized range) and exponential moving averaged weights. For data augmentation, techniques such as rotate90 covering all four orientations and low-resolution noise injection (for the super-resolution task) are also used and found to contribute to better model outputs. To eliminate noise in the total precipitation maps, a minimum filter as post-processing is also utilized.

\subsection{Evaluation Strategies}
To effectively assess the three cascaded diffusion models mentioned in this work, the two evaluation strategies are undertaken. First, quantitative metrics involving MAE, PSNR, SSIM and FID scores are used to assess the one-step performance over all the epochs. And second, rollout analysis using SSIM evaluation over the forecast generated in an auto-regressive manner starting with an initial IR 10.8{\textmu m} assisted with forecasted ERA5 data is performed over the entire cyclone duration.

\section{Results}

Performance evaluation of the best performing model checkpoint over four distinct metrics is shown in Table \ref{tbl:performance-metrics}. These results underscore the remarkable predictive capabilities of all three diffusion models for forecasting purposes, consistently surpassing the thresholds of 20dB and 0.5 for PSNR and SSIM values, respectively. Additionally, the MAE (measured over normalised images) is found to consistently yield values below 0.25, while the FID scores remain below 1 for all three models.

\begin{table}[!h]
\small
\centering
\vspace{-0.5em}
\caption{Task-wise performance metrics over the entire test set}
\label{tbl:performance-metrics}
\begin{tabular}{lrrrrrrr}
\toprule
\textbf{Task} & \textbf{Best Epoch} & \textbf{MAE $\downarrow$} & \textbf{PSNR $\uparrow$} & \textbf{SSIM $\uparrow$} & \textbf{FID $\downarrow$} \\
\midrule
Forecasting & 180 & 0.200 & 24.359 & 0.572 & 0.168 \\
Super Resolution & 240 & 0.186 & 25.760 & 0.658 & 0.214 \\
Precipitation Modelling & 260 & 0.134 & 25.943 & 0.643 & 0.907 \\
\bottomrule
\end{tabular}
\end{table}

\begin{figure}[!h]
\centering
\includegraphics[width=0.8\hsize]{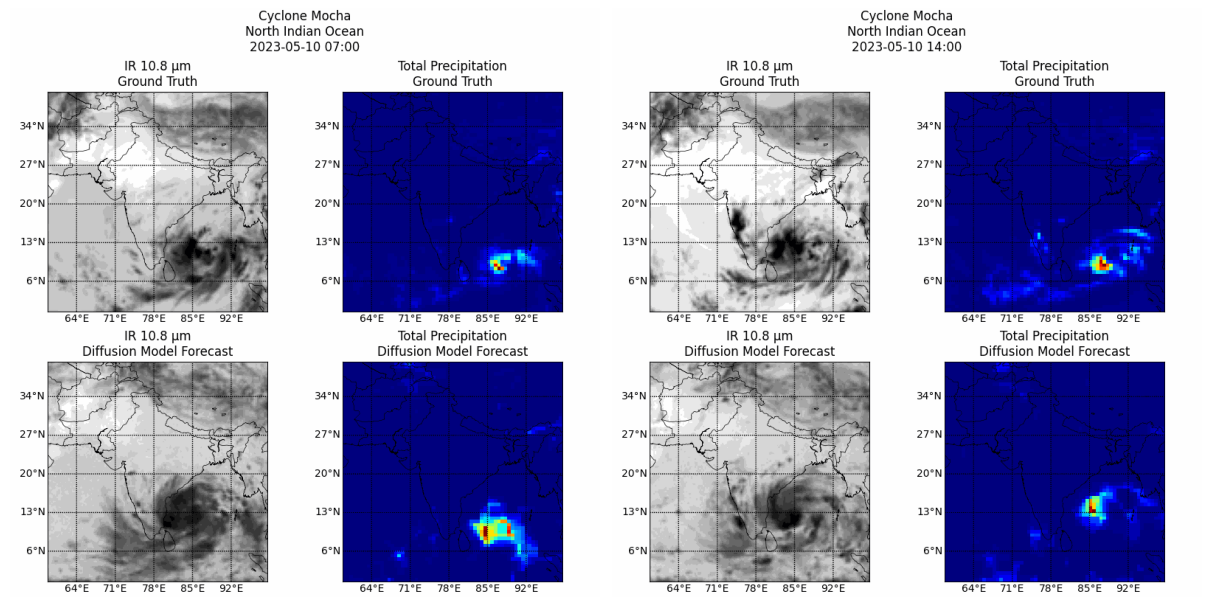}
\caption{Forecast at 31h (left) and 38h (right) of Cyclone Mocha over the North Indian Ocean on 10th May 2023. The upper rows resemble the ground truth IR 10.8{\textmu m} satellite image and total precipitation while the lower rows show the forecast generated at that particular timestep.}
\label{fig:tc-forecast}
\end{figure}

For forecasts of all cyclones belonging to the region wise test set in Table~\ref{tbl:cyclone-basins} (such as the one for Cyclone Mocha shown in Fig.~\ref{fig:tc-forecast}), upon closer examination of the SSIM charts generated over the entire cyclone duration as displayed in Fig.~\ref{fig:forecast-horizon}, a notable decline can be observed in the majority of cyclone forecasts around the 36-hour mark. Given the challenge of identifying an absolute rollout with certainty, the consistent occurrence of sharp "dips" at approximately the 36-hour mark (approx. 30 minutes on an Nvidia A30/RTX 2080Ti) implies that such a time length can be considered a reliable horizon where the generated forecast can be estimated to closely align with the actual conditions.

\section{Conclusion}

This work presents a novel cascaded diffusion model architecture for forecasting tropical cyclones supported by using a custom-built data processing pipeline and trained on IR 10.8{\textmu m} in addition to ERA5 atmospheric reanalysis data. With strong enough capabilities for forecasts with horizons of around 36 hours, when integrated with atmospheric data, like ERA5, these instances of cost-effective AI-based modelling, optimised for single GPUs, facilitate affordable, almost real-time, precise, and photorealistic forecasts. This makes them particularly suitable for highly vulnerable regions facing critical forecasting demands but are financially constrained. Future iterations of this work aim to explore the modeling of cyclones over extended periods (eg. 10 years) and incorporate a broader array of atmospheric data variables. Leveraging the recent development in machine learning for ERA5 forecasting~\cite{Scholz.2023}~\cite{Bonev.2023} and the use of more powerful computational platforms, including multi-GPU Nvidia A100 setups, we aim to enhance the model's predictive accuracy and efficiency over limited execution timeframes.

\bibliographystyle{vancouver}
\bibliography{bibliography}

\clearpage

\appendix
\renewcommand\thesection{Appendix \Alph{section}}
\renewcommand\thesubsection{\Alph{section}.\arabic{subsection}} 
\renewcommand\thetable{\Alph{section}.\arabic{table}}  
\renewcommand\thefigure{\Alph{section}.\arabic{figure}}  
\setcounter{table}{0}

\section{Additional Data Description}
\subsection{ERA5 Variables}

\begin{table}[!h]
\small
\centering
\vspace{-1em}
\caption{List of ERA5 variables considered for condition data}
\label{tbl:era5-data}
\vspace{0.75em}
\begin{tabular}{llp{8cm}}
\toprule
\textbf{Name} & \textbf{Unit} \\ \midrule
10m u-component of wind & ms$^{-1}$ \\
10m v-component of wind & ms$^{-1}$ \\
Total cloud cover & - \\
Total precipitation & m \\ \bottomrule
\end{tabular}
\end{table}

For the experiments performed, due to the historical nature of the cyclones, forecasted ERA5 refers to the ERA5 reanalysis data generated by ECMWF. In deployment scenarios, due to ERA5 having a lag of 5 days, actual forecasted atmospheric data (corresponding to the three variables) from forecasting systems can be used. Therefore, to ensure consistency with common forecasting systems such as Global Forecast System (GFS), these specific four variables are chosen.

\section{Additional Results}
\subsection{Forecast SSIM}

\setcounter{figure}{0} 
\begin{figure}[!h]
\centering
\begin{subfigure}[b]{0.32\textwidth}
     \centering
     \captionsetup{justification=centering}
     \includegraphics[width=\textwidth]{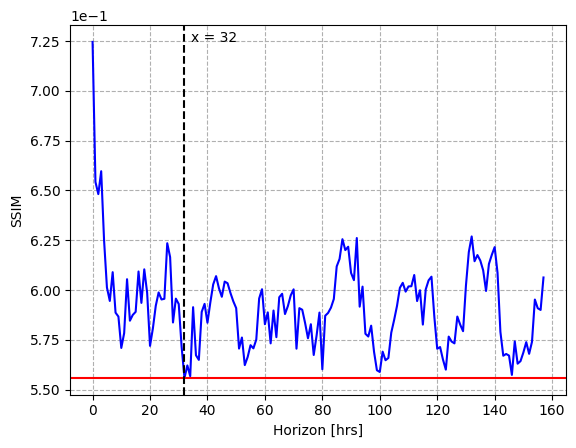}
     \subcaption{Mocha\\(North Indian Ocean)}
\end{subfigure}
\hfill
\begin{subfigure}[b]{0.32\textwidth}
     \centering
     \captionsetup{justification=centering}
     \includegraphics[width=\textwidth]{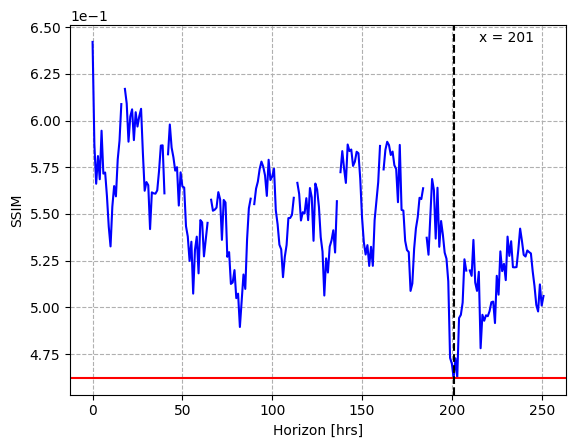}
     \subcaption{Ida\\(North Atlantic Ocean)}
\end{subfigure}
\hfill
\begin{subfigure}[b]{0.32\textwidth}
     \centering
     \captionsetup{justification=centering}
     \includegraphics[width=\textwidth]{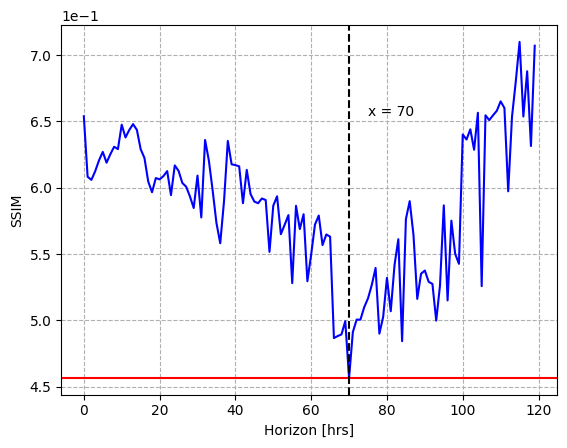}
     \subcaption{Rosyln\\(Eastern Pacific Ocean)}
\end{subfigure}

\begin{subfigure}[b]{0.32\textwidth}
     \centering
     \captionsetup{justification=centering}
     \includegraphics[width=\textwidth]{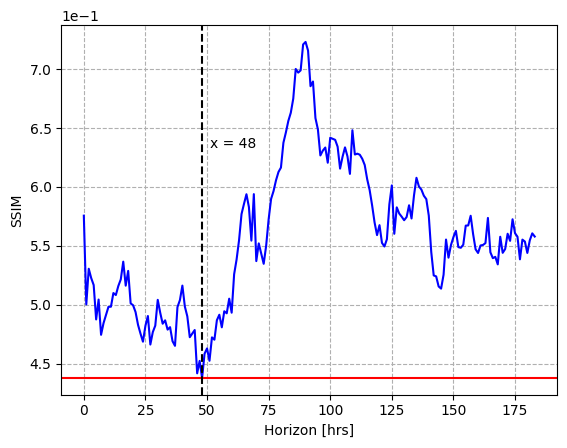}
     \subcaption{Molave\\(Western Pacific Ocean)}
\end{subfigure}
\hfill
\begin{subfigure}[b]{0.32\textwidth}
     \centering
     \captionsetup{justification=centering}
     \includegraphics[width=\textwidth]{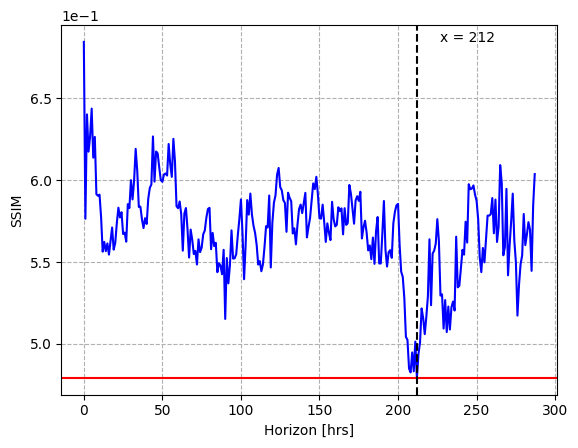}
     \subcaption{Gombe\\(SW Indian Ocean)}
\end{subfigure}
\hfill
\begin{subfigure}[b]{0.32\textwidth}
     \centering
     \captionsetup{justification=centering}
     \includegraphics[width=\textwidth]{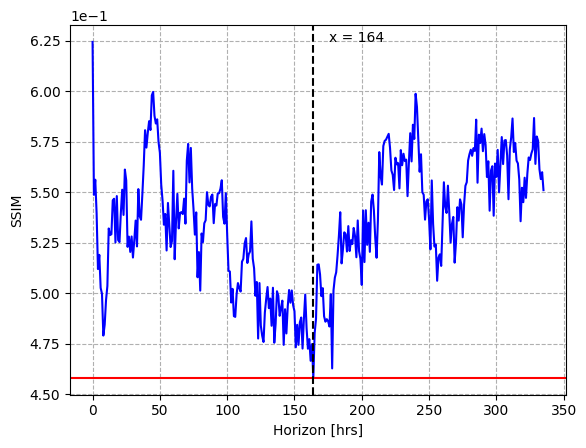}
     \subcaption{Veronica\\(Australia)}
\end{subfigure}
\caption{SSIM values over the entire cyclonic duration. The dashed lines indicate the hourly marks at which the minimum SSIM values are obtained for each cyclone.}
\label{fig:forecast-horizon}
\end{figure}

\end{document}